\documentclass[10pt, conference, a4paper]{IEEEtran}
\usepackage{amsmath, amsfonts,  amssymb, color, textcomp}
\usepackage[all]{xy}
\usepackage{graphicx}
\usepackage{color}
\usepackage{latexsym}
\usepackage{bm}
\usepackage{url}
 \usepackage[pass]{geometry}
    
\newcommand{\btb}{\mathbf{B}_{tb}^{(\lambda)}}
\newcommand{\bterm}{\mathbf{B}_{[0,L-1]}}
\newcommand{\binf}{\mathbf{B}_{[0,\infty]}}

\newcommand\scalemath[2]{\scalebox{#1}{\mbox{\ensuremath{\displaystyle #2}}}}

\DeclareMathOperator{\Supp}{Supp}

\long\def\symbolfootnote[#1]#2{\begingroup%
\def\thefootnote{\fnsymbol{footnote}}\footnote[#1]{#2}\endgroup} 
\begin{document}

% paper title
\title{Free Pseudodistance Growth Rates for Spatially Coupled LDPC Codes over the BEC}
\author{
\authorblockN{Cunlu Zhou$^\dag$, David G. M. Mitchell$^*$, and Roxana Smarandache$^\dag$}
\authorblockA{$^\dag$Department of Mathematics, University of Notre Dame, Notre Dame, Indiana, USA\\
\{czhou3, rsmarand\}@nd.edu\\
$^*$Klipsch School of Electrical and Computer Engineering, New Mexico State University, Las Cruces, New Mexico, USA\\
dgmm@nmsu.edu}\vspace{0mm}}

\maketitle
\begin{abstract}
The minimum pseudoweight is an important parameter related to the decoding performance of LDPC codes with iterative message-passing decoding. In this paper, we consider ensembles of periodically time-varying spatially coupled LDPC (SC-LDPC) codes and the pseudocodewords arising from their finite graph covers of a fixed degree. We show that for certain $(J,K)$-regular SC-LDPC code ensembles and a fixed cover degree, the typical minimum pseudoweight of the unterminated (and associated tail-biting/terminated) SC-LDPC code ensembles grows linearly with the constraint (block) length as the constraint (block) length tends to infinity. We prove that one can bound the the free pseudodistance growth rate over a BEC from below (respectively, above) using the associated tail-biting (terminated) SC-LDPC code ensemble and show empirically that these bounds coincide for a sufficiently large period, which gives the exact free pseudodistance growth rate for the SC-LDPC ensemble considered. 
\end{abstract}

\section{Introduction}
Pseudocodewords have been shown to play a key role in understanding the decoding performance of low-density parity-check (LDPC) codes with message-passing iterative decoders \cite{kv03,vk07,ks07}. In \cite{kv03} and \cite{vk07}, it was shown that an iterative decoder cannot distinguish between the original Tanner graph and any of its finite graph covers. As a consequence, the performance of iterative decoders is characterized by the pseudocodewords associated with all of the finite covers. In particular, the \textit{minimum pseudoweight} (or \textit{pseudodistance}) plays, in iterative decoding, the role that the minimum distance does for maximum likelihood (ML) decoding \cite{fkkr01, vk07, sv07}. For certain protograph-based LDPC code ensembles, it has been shown in \cite{sdw11} that the minimum pseudoweight, typical of most ensemble members, obtained from graph covers for a fixed degree grows linearly with the block length $n$ as $n\to \infty$. A large \textit{pseudodistance growth rate} (or \textit{typical relative minimum pseudoweight}) means that, asymptotically, most pseudocodewords from the ensemble are \textit{``good pseudocodewords''}. 

Spatially coupled LDPC (SC-LDPC), or LDPC convolutional (LDPCC), codes \cite{fz99} are constructed by \emph{coupling} together a sequence of $L$ uncoupled (or disjoint) Tanner graphs into a single coupled chain, thus introducing memory into the encoding process. SC-LDPC codes have been shown to have excellent iterative decoding {thresholds} \cite{lscz10,kru13} and good asymptotic minimum distance properties \cite{mlc15,mplc11}. In \cite{mplc11} and \cite{mpc13}, Mitchell et al. showed how to bound the \textit{free distance growth rate} of an SC-LDPC code ensemble from above and below, resulting in an exact free distance growth rate of the code ensemble. In \cite{spv09}, Smarandache et al. studied the pseudocodeword problem from the perspective of convolutional codes. They proved that for a class of quasi-cyclic (QC) based time-invariant LDPCC codes \cite{tss04}, the minimum pseudoweight of an LDPCC code is lower bounded by the minimum pseudoweight of its ``wrapped" QC code.

%Spatially coupled LDPC (SC-LDPC) codes are constructed by \emph{coupling} together a series of $L$ disjoint (or uncoupled) Tanner graphs into a single coupled chain. They can be viewed as a type of LDPC convolutional (LDPCC) code \cite{fz99}, since spatial coupling is equivalent to introducing memory into the encoding process. SC-LDPC codes have been shown to have excellent iterative decoding {thresholds} \cite{lscz10,kru13}, good asymptotic minimum distance properties and trapping set properties \cite{mlc13,mplc11}. Moreover, it has been proven analytically for general memoryless binary-input symmetric-output (MBS) channels that for a class of $(J,K)$-regular SC-LDPC code ensembles the belief propagation (BP) decoding thresholds achieve the maximum a posteriori probability (MAP) decoding thresholds of the underlying $(J,K)$-regular LDPC block code ensembles, a phenomenon called \emph{threshold saturation} \cite{kru13}. In \cite{mplc11} and \cite{mpc13}, Mitchell et al. showed how to bound the \textit{free distance growth rate} of an SC-LDPC code ensemble from above and below, resulting in an exact free distance growth rate of the code ensemble. In \cite{spv09}, Smarandache et al. studied the pseudocodeword problem from the perspective of convolutional codes. They proved that for a class of quasi-cyclic (QC) based time-invariant LDPCC codes (see \cite{tss04}), the minimum pseudoweight of an LDPCC code is lower bounded by the minimum pseudoweight of its ``wrapped" QC code. 
 
In this paper, we consider ensembles of protograph-based periodically time-varying SC-LDPC codes and their resulting pseudocodewords obtained as projections of codewords from their finite-degree graph covers. We show that for certain $(J,K)$-regular SC-LDPC code ensembles, the typical minimum pseudoweight obtained from graph covers for a fixed degree of the unterminated (and associated tail-biting/terminated) SC-LDPC code ensembles grows linearly with the constraint (block) length as the constraint (respectively, block) length tends to infinity. We prove that a similar approach to that from \cite{mplc11} and \cite{mpc13} can be used to obtain the exact \textit{free pseudodistance growth rate} of the periodically time-varying SC-LDPC code ensembles over a binary erasure channel (BEC). More specifically, we first prove that, on average, the ensemble free pseudodistance can be bounded below by the pseudodistance of an associated tail-biting ensemble and above by the pseudodistance of an associated terminated ensemble, and we derive the upper and lower bounds for the free pseudodistance growth rate of the ensemble. 

To demonstrate empirically these theoretical analyses, we perform numerical experiments for degree-$2$ and degree-$3$ graph covers.\footnote{We limit our consideration in this paper to degree-2 and degree-3 covers due to the high computational complexity required to evaluate graph covers of larger degrees.} Besides obtaining the aforementioned bounds, we show that these bounds coincide for a sufficiently large period thus give the exact free pseudodistance growth rate of the ensemble considered. We observe that the free pseudodistance growth rate of the unterminated $(J,K)$-regular SC-LDPC code ensemble is much larger than the underlying $(J,K)$-regular LDPC code ensemble. Also, by comparing to the results in \cite{mplc11}, we find that the free pseudodistance growth rate is smaller than the free distance growth rate, as expected.\footnote{Note that this paper analyzes the \textit{free pseudodistance growth rate} which is an important indicator of the decoding performance of iterative decoding, while in \cite{mplc11}, the \textit{free distance growth rate} is derived mainly as a performance indicator for ML decoding.} 
%Secondly, the analysis and computation for the pseudo case is more complicated than the non-pseudo case, especially the computation. Structures of the optimization problem involved have been taken into account to choose the right optimization solvers and algorithms, without which the computation is significantly more difficult.

The paper is structured as follows. In Section \ref{sec:bkgd}, we describe the necessary background including the protograph construction method, graph-cover pseudocodewords, and convolutional protographs including a discussion of two different ways of terminating SC-LDPC codes which will be used to obtain lower and upper bounds in the following section. In Section \ref{sec:psdist}, we conduct the free pseudodistance analysis of SC-LDPC code ensembles with finite-degree covers over a BEC. We first prove bounds for the ensemble average free pseudodistance in Section \ref{sec:pdbd} and then derive related bounds for the free pseudodistance growth rates of the code ensembles considered in Section \ref{sec:pdgrowth}. Numerical results for the pseudodistance growth rates of a $(3,6)$-regular SC-LDPC code ensemble are presented in Section \ref{sec:num}. Finally, concluding remarks are given in Section \ref{sec:conc}.

\section{Background}\label{sec:bkgd}
%A protograph \cite{tho03} is a small bipartite graph that is used to derive a larger graph by taking an $N$-fold graph cover \cite{mas77}, or ``lifting'', of the protograph. It is an important feature of this construction that each lifted code inherits the degree distribution and graph neigbourhood structure of the protograph.
A protograph \cite{tho03} is a small bipartite graph that is used to derive a larger graph by ``lifting'', i.e., taking an $N$-fold graph cover of the protograph. The lifted graph preserves the graph neigbourhood structure and degree distribution of the protograph. The protograph can be represented by a \emph{base} $b_c \times b_v$ biadjacency matrix $\mathbf{B}=[b_{x,y}]$, where $b_{x,y}$, $1\leq x \leq b_c$, $1\leq y \leq b_v$, is the number of edges connecting variable node $v_y$ to check node $c_x$. The parity-check matrix $\mathbf{H}$ of a protograph-based LDPC block code can be constructed by replacing each non-zero entry in $\mathbf{B}$  by a sum of $ b_{x,y}$  permutation matrices of size $N\times N$ and each zero entry by the $N\times N$ all-zero matrix. The ensemble of protograph-based LDPC block codes with block length $n = N n_v$  is defined by the set of matrices $\mathbf{H}$ that can be derived from a given protograph by choosing all possible combinations of $N \times N$ permutation matrices.

\subsection{Graph-Cover Pseudocodewords}\label{sec:pscw}
Let $m$ be an integer. Given a Tanner graph $G$ with $n$ variable nodes, consider an $m$-fold graph cover of $G$, denoted as $G^m$. Let $c=(c_{1,1},\ldots,c_{1,m},\ldots, c_{n,1},\ldots, c_{n,m})$ be a codeword of $G^m$, then $\mathbf{w}=[w_1,\ldots, w_n]$ is a \textit{pseudocodeword} of $G$, where $w_i=\sum_{k=1}^m c_{i,k}$, $i=1,\ldots,n$ \cite{sdw11}. The \textit{pseudoweight} of $\mathbf{w}$ over a BEC is $|\Supp(\mathbf{w})|$, the number of nonzeros in $\mathbf{w}$, denoted as $p(\mathbf{w})$. The \textit{pseudodistance}, $w_{min}^m$, for finite covers of a fixed degree $m$ of $G$ is defined as the minimum pseudoweight among all non-zero pseudocodewords from the degree-$m$ covers. Without danger of ambiguity, we will use $w_{min}$ instead; however, it should be emphasized that in our paper $w_{min}$ is not defined for all possible finite-degree covers of $G$. In addition, we will use the term \textit{pseudodistance} and \textit{minimum pseudoweight} interchangeably. %Since in this paper all pseudodistance and pseudoweight quantities are obtained from finite covers of a fixed degree $m$, we will simply use $w_{min}$ instead; however, it should be understood that these terms do not reflect the quantities over all possible finite-degree covers.%We will use the term pseudodistance and minimum pseudoweight interchangeably throughout the paper. %Note that we define the \textit{pseudodistance} only for the finite covers of a fixed degree-$m$ not for all finite-degree covers of $G$.%\footnote{Note that, in this paper, all pseudodistance and pseudoweight quantities are obtained from a particular degree-$m$ cover. As such, we do not include the cover degree $m$ in the notation for clarity; however, it should be understood that these terms do not reflect the quantities over all possible degree-$m$ covers.}%In addition, to avoid confusion, we will use the term degree-$m$ cover instead of $m$-fold cover (although they essentially mean the same) when we consider pseudocodewords. 

\subsection{Convolutional protographs}\label{sec:convproto}
An ensemble of unterminated SC-LDPC codes can be described by a {\em convolutional protograph} \cite{mlc15} with base matrix
\begin{equation}\label{convbase}
\scalemath{1.0}{
\mathbf{B}_{[0,\infty]}=\left[
\begin{array}{cccccc}
\mathbf{B}_{0} &   & \\
\mathbf{B}_{1} &\mathbf{B}_{0} &  \vspace{-2.2mm}\\
\vdots &\mathbf{B}_{1} & \vspace{-2mm}\ddots \\
\mathbf{B}_{m_s} & \vdots& \ddots\\
 & \mathbf{B}_{m_s}& \vspace{-2mm}\\
 & & \ddots\\
\end{array}\right]},
\end{equation}
where $m_s$ denotes the \textit{syndrome former memory} of the convolutional codes and the $b_c \times b_v$ {\em component base matrices} $\mathbf{B}_{i}$, $i=0,\dots,m_s$, represent the edge connections from the $b_v$ variable nodes at time $t$ to the $b_c$ check nodes at time $t+i$. An ensemble of time-varying SC-LDPC codes can then be formed from $\mathbf{B}_{[0,\infty]}$ using the protograph construction method  described above, resulting in the associated parity-check matrix\vspace{0mm}\\
\mbox{\scriptsize{$\mathbf{H}_{[0,\infty]}=$}}\vspace{0.5mm}\\
\hspace*{1mm}\scalebox{0.9}{\mbox{\scriptsize{$
\left[ \begin{array}{cccccc}
\mathbf{H}_{0}(0) & & & & \\
\mathbf{H}_{1}(1) & \mathbf{H}_{0}(1)& & & \\
\vdots & \vdots& & \ddots& \\
\mathbf{H}_{m_s}(m_s) & \mathbf{H}_{m_s-1}(m_s)&\cdots &\mathbf{H}_{0}(m_s) & \\
& \mathbf{H}_{m_s}(m_s+1) & \mathbf{H}_{m_s-1}(m_s+1)&\cdots
&\mathbf{H}_{0}(m_s+1) \\
&\ddots&\ddots&&\ddots
\end{array} \right].$}
}}\vspace{1mm}
\noindent A rate $R=1-Nb_c/Nb_v=1-b_c/b_v$ time-varying SC-LDPC code with parity-check matrix $\mathbf{H}_{[0,\infty]}$ is periodically time-varying with period $T$ if $\mathbf{H}_{i}(t)$ is periodic, i.e., $\mathbf{H}_{i}(t)=\mathbf{H}_{i}(t+T), \forall ~i,t$, and if $\mathbf{H}_{i}(t)=\mathbf{H}_{i}, \forall ~i,t$, the code is \emph{time-invariant}. We call $\nu_s = N(m_s+1)b_v$ the \emph{decoding constraint length}.

Starting from the base matrix $\mathbf{B}$ of a block code ensemble, one can construct SC-LDPC code ensembles with the same computation trees. This is achieved by an {\em edge spreading} procedure (see \cite{mlc15} for details) that divides the edges from each variable node in the base matrix $\mathbf{B}$ among $m_s+1$ component base matrices $\mathbf{B}_i$, $i=0,\dots,m_s$, such that the condition $\mathbf{B}_0+\mathbf{B}_1+\cdots+\mathbf{B}_{m_s}=\mathbf{B}$ is satisfied. For example, a (3,6)-regular SC-LDPC ensemble with $m_s=2$ can be formed from the block base matrix $\mathbf{B} = [\hspace{1mm}3\hspace{2mm} 3\hspace{1mm}]$ by defining the component base matrices
$\mathbf{B}_0=[\hspace{1mm}1\hspace{2mm} 1\hspace{1mm}]=\mathbf{B}_1=\mathbf{B}_2\hspace{0.5mm}$.

From a convolutional protograph with base matrix $\mathbf{B}_{[0,\infty]}$, we can form a periodically time-varying $N$-fold graph cover with period $T$ by choosing, for the $b_c\times b_v$ submatrices $\mathbf{B}_0,\mathbf{B}_1,\ldots,\mathbf{B}_{m_s}$ in the first $T$ columns of $\mathbf{B}_{[0,\infty]}$, a set of $N\times N$ permutation matrices randomly and independently to form $Nb_c \times Nb_v$ submatrices $\mathbf{H}_0(t),\mathbf{H}_1(t+1),\ldots,\mathbf{H}_{m_s}(t+m_s)$, respectively, for $t=0,1,\ldots,T-1$. These submatrices are then repeated periodically (indefinitely) to form $\mathbf{H}_{[0,\infty]}$ such that $\mathbf{H}_i(t+T)=\mathbf{H}_i(t)$, $\forall i,t$. An ensemble of periodically time-varying SC-LDPC codes with period $T$, rate $R=1-Nb_c/Nb_v=1-b_c/b_v$, and decoding constraint length $\nu_s=N(m_s+1)b_v$ can then be derived by letting the permutation matrices used to form $\mathbf{H}_0(t),\mathbf{H}_1(t+1),\ldots,\mathbf{H}_{m_s}(t+m_s)$, for $t=0,1,\ldots,T-1$, vary over the $N!$ choices of permutation matrix.
 
\subsection{Termination of SC-LDPC codes}\label{sec:termination}
Suppose that we start the convolutional code with parity-check matrix defined in $(\ref{convbase})$ at time $t=0$ and terminate it after $L$ time instants. The resulting finite-length base matrix is then given by
\begin{equation}\label{termbase}
\scalemath{1.0}{
\mathbf{B}_{[0,L-1]}=\left[
\begin{array}{ccc}
\mathbf{B}_0 & &\\
%\mathbf{B}_1 & \ddots & \mathbf{B}_1\\
\vdots & \ddots &  \\
\mathbf{B}_{m_s} &  & \mathbf{B}_0 \\
& \ddots & \vdots\\
& & \mathbf{B}_{m_s}
\end{array}\right]_{(L+m_s)b_c \times Lb_v}}.
\end{equation}
The matrix $\mathbf{B}_{[0,L-1]}$ can be considered as the base matrix of a terminated protograph-based SC-LDPC code ensemble. Termination in this fashion results in a rate loss. The design rate of the terminated code ensemble is given as
\begin{equation}\label{termrate}
    R_L=1-\left(\frac{L+m_s}{L}\right)\frac{b_c}{b_v}=1-\left(\frac{L+m_s}{L}\right)\left(1-R\right),
\end{equation}
where $R=1-Nb_c/Nb_v=1-b_c/b_v$ is the rate of the unterminated convolutional code ensemble. Note that, as the \emph{termination factor} $L$ increases, the rate increases monotonically and approaches the rate of the unterminated convolutional code ensemble.

The convolutional base matrix $\mathbf{B}_{[0,\infty]}$ can also be terminated using \emph{tail-biting} \cite{st79, mw86}. Here, for any $\lambda \geq m_s$, the last $b_cm_s$ rows of the terminated parity-check matrix $\mathbf{B}_{[0,\lambda-1]}$ are removed and added to the first $b_cm_s$ rows to form the $\lambda b_c \times \lambda b_v$ tail-biting parity-check matrix $\mathbf{B}_{tb}^{(\lambda)}$ with tail-biting termination factor $\lambda$. 
\noindent  %Note that, if $m_s=1$ and $\lambda = 1$, the tail-biting base matrix is simply the original block base matrix, i.e., $\mathbf{B}_{tb}^{(1)}=\mathbf{B}$. 
Terminating $\mathbf{B}_{[0,\infty]}$ in such a way preserves the design rate of the ensemble, i.e., $R_\lambda =1-\lambda b_c/\lambda b_v=1-b_c/b_v=R$, and we see that $\mathbf{B}_{tb}^{(\lambda)}$ has exactly the same degree distribution as the original block base matrix $\mathbf{B}$.

\section{Free pseudodistance analysis of\\ SC-LDPC code ensembles with \\finite-degree covers over the BEC}\label{sec:psdist}
In this section, we investigate the free pseudodistance of periodically time-varying SC-LDPC code ensembles with finite-degree covers over a BEC by deriving bounds for the ensemble average free pseudodistance using terminated and tail-biting SC-LDPC code ensembles.

\subsection{Free pseudodistance bounds for SC-LDPC code ensembles with degree-m covers}\label{sec:pdbd}
Let $E(T)$ denote the ensemble of unterminated periodically time-varying SC-LDPC codes as described in Section \ref{sec:convproto}. Let $E_{tb}(\lambda)$ denote the associated ensemble of tail-biting SC-LDPC codes derived from the base matrix $\mathbf{B}_{tb}^{(\lambda)}$ with termination factor $\lambda=T$, referred to simply as the \emph{tail-biting ensemble}. Let $E_{t}(L)$ denote the associated ensemble of terminated SC-LDPC codes derived from the base matrix $\mathbf{B}_{[0,L-1]}$ with block length $n=LNb_v$ and termination factor $L=T$, referred to as the \emph{terminated ensemble}. For a fixed integer $m$, consider the degree-$m$ graph covers of a code ensemble, i.e., for each code in the ensemble, consider all of its degree-$m$ covers. We define the \emph{ensemble average minimum pseudoweight} over all of the pseudocodewords from all of the degree-$m$ covers of all of the codes in the ensemble. Let $\bar{w}_{free}(T)$, $\bar{w}_{min,tb}(\lambda)$ and $\bar{w}_{min,t}(L)$ denote the ensemble average pseudodistance of $E(T)$, $E_{tb}(\lambda)$, and $E_{t}(L)$, respectively. 
\vspace{1mm}
\newtheorem{tlbt}{Lemma}
\begin{tlbt}\label{tlbt}
Let $C$ be an arbitrary SC-LDPC code drawn from ensemble $E(T)$ and consider a degree-$m$ cover $C^m$ of $C$. Let $C_{tb}(\lambda)$ and $C_{tb}^m(\lambda m)$ be the associated tail-biting codes of $C$ and $C^m$, respectively, with tail-biting termination factor $\lambda$, $\lambda\in \{T,2T,3T,\ldots\}$, $T\geq m_s+1$.\footnote{Note that we must select a multiple of the period $T$ as the termination factor so that the wrapped word is a codeword in the tail-biting code. For more details, see \cite{mpc13}.} Let
$\mathbf{w}=[w_1,w_2,\ldots]$ be an arbitrary pseudocodeword of $C$ obtained from a degree-$m$ cover, where $w_i=\sum_{k=1}^m c_{i,k}$, $i=1,2,\ldots$, and $\mathbf{c}=(c_{1,1},\ldots,c_{1,m}, c_{2,1},\ldots,c_{2,m},\ldots)$ is a codeword of $C^m$. Then the ``wrapped'' vector $\hat{\mathbf{w}}=[\hat{w}_1, \hat{w}_2,\ldots,\hat{w}_{\lambda Nb_v}]$, where $\hat{w}_i=\sum_{k=1}^m(\sum_{j=0}^{\infty}c_{i+j\lambda Nb_v,k} \bmod 2)$, $i=1,2,\ldots,\lambda Nb_v$, is a pseudocodeword of $C_{tb}(\lambda)$ obtained from a degree-$m$ cover. Furthermore, we have pseudoweight $p(\hat{\mathbf{w}})\leq p(\mathbf{w})$ over a BEC.
\end{tlbt}
\vspace{1mm}
\emph{Sketch of Proof}. Following the argument in \cite{mpc13}, given a codeword $\mathbf{c}$ in $C^m$, the wrapped vector %$\hat{\mathbf{c}}= (\sum_{j=0}^{\infty}c_{1+j\lambda Nb_v,1} \mod 2, \ldots, \sum_{j=0}^{\infty}c_{1+j\lambda Nb_v,m} \mod 2, \ldots,\linebreak \sum_{j=0}^{\infty}c_{\lambda Nb_v+j\lambda Nb_v,1} \mod 2, \ldots, \sum_{j=0}^{\infty}c_{\lambda Nb_v+j\lambda Nb_v,m} \mod 2)$ 
$\hat{\mathbf{c}}= (\sum_{j=0}^{\infty}c_{1+j\lambda Nb_v,1}, \ldots, \sum_{j=0}^{\infty}c_{1+j\lambda Nb_v,m}, \ldots,\linebreak \sum_{j=0}^{\infty}c_{\lambda Nb_v+j\lambda Nb_v,1}, \ldots, \sum_{j=0}^{\infty}c_{\lambda Nb_v+j\lambda Nb_v,m}),$ where all sums are performed modulo 2, is a codeword in $C_{tb}^m(\lambda m)$. By summing every $m$ entries in $\hat{\mathbf{c}}$, we obtain $\hat{\mathbf{w}}$, a pseudocodeword from a degree-$m$ cover of $C_{tb}^{(\lambda)}$. Clearly, $|\Supp(\hat{\mathbf{w}})|\leq |\Supp(\mathbf{w})|$, i.e., over a BEC, the pseudoweight $p(\hat{\mathbf{w}})\leq p(\mathbf{w})$. \hfill $\Box$
\vspace{1mm}
\newtheorem{example}{Example}
\begin{example}
To illustrate the idea in Lemma \ref{tlbt}, consider an ensemble $E$  of time-invariant SC-LDPC codes constructed from the block base matrix 
$\mathbf{B}=[\hspace{1mm}2\hspace{2mm} 2\hspace{1mm}]$ with component base matrices $\mathbf{B}_0 = \mathbf{B}_1 = [\hspace{1mm}1\hspace{2mm} 1\hspace{1mm}]$.\footnote{Note that we drop the notation of period $T$ for time invariant codes.} Then we have the base matrix of the convolutional protograph

\begin{equation*}
\mathbf{B}_{[0,\infty]}=\left[
\begin{array}{cccccc}
1\hspace{2mm} 1 &   &  &\\
1\hspace{2mm} 1 & 1\hspace{2mm} 1 &  &\\
 & 1\hspace{2mm} 1 & 1\hspace{2mm} 1 & \vspace{-2mm}\\
 & & 1\hspace{2mm} 1 & \ddots \vspace{-2mm}\\
 & & & \ddots\\
\end{array}\right].
\end{equation*}

\noindent For the purpose of illustration, let's consider the trivial ensemble with $1$-fold cover, so $\binf = E$. Consider pseudocodewords from a degree-$2$ cover $\mathbf{P}_{[0,\infty]}$ of $\binf$, 

\begin{equation*}
\mathbf{P}_{[0,\infty]}=\left[
\begin{array}{cccccc}
\mathbf{I}_2\hspace{2mm}\mathbf{I}_2 &   &  &\\
\mathbf{I}_2\hspace{2mm}\mathbf{I}'_2 & \mathbf{I}_2\hspace{2mm}\mathbf{I}_2 &  & \\
 & \mathbf{I}_2\hspace{2mm}\mathbf{I}'_2 & \mathbf{I}_2\hspace{2mm}\mathbf{I}_2 & \vspace{-2mm}\\
 & & \mathbf{I}_2\hspace{2mm}\mathbf{I}'_2 & \ddots \vspace{-2mm}\\
 & & & \ddots\\
\end{array}\right],
\end{equation*}
where $\mathbf{I}_2 = \left[\begin{smallmatrix} 1&0\\ 0&1 \end{smallmatrix}\right]$ and $\mathbf{I}'_2 = \left[\begin{smallmatrix} 0&1\\ 1&0 \end{smallmatrix}\right]$. When the tail-biting termination factor $\lambda=2$, we have

\begin{equation*}\mathbf{B}_{tb}^{(\lambda)}=\mathbf{B}_{tb}^{(2)}=\left[
\begin{array}{cc}
1\hspace{2mm} 1 & 1\hspace{2mm} 1 \\
1\hspace{2mm} 1 & 1\hspace{2mm} 1 \\
\end{array}\right]\end{equation*}
and
\begin{equation*}\mathbf{P}_{tb}^{(\lambda)}=\mathbf{P}_{tb}^{(2)}=\left[
\begin{array}{cc}
\mathbf{I}_2\hspace{2mm}\mathbf{I}_2 & \mathbf{I}_2\hspace{2mm}\mathbf{I}'_2 \\
\mathbf{I}_2\hspace{2mm}\mathbf{I}'_2 & \mathbf{I}_2\hspace{2mm}\mathbf{I}_2 \\
\end{array}\right].\end{equation*}

Here, $m=2$ and $\binf$ defines $C$, $\mathbf{P}_{[0,\infty]}$ defines $C^m$, $\mathbf{B}_{tb}^{(\lambda)}$ defines $C_{tb}(\lambda)$, and $\mathbf{P}_{tb}^{(\lambda)}$ defines $C_{tb}^m(\lambda m)$ in Lemma \ref{tlbt}. Consider a $2$-cover pseudocodeword of $C$ which is constructed by summing every two bits of a codeword (in general not unique) of the code $C^m$, e.g., $\mathbf{w}=[1,1,2,0,1,1,1,1,1,1,0,\ldots]$ is constructed from $\mathbf{c}=(c_{1,1},c_{1,2},\ldots,c_{10,1},c_{10,2},0,\ldots)=([1,0],[1,0],[1,1],[0,0],[0,1],[1,0],[0,1],[0,1],[0,1],[1,0],$ $0,\ldots)\in C^2$. Note that we grouped the associated two bits together for easy interpretation. Since $\lambda N b_v = 2\times 1 \times 2 = 4$, by ``wrapping'' the codeword $\mathbf{c}$, we obtain a vector of length $8$ ($=4m$)  $\hat{\mathbf{c}}=(\hat{c}_{1,1},\hat{c}_{1,2},\ldots,\hat{c}_{4,1},\hat{c}_{4,2})=([1,0],[1,0],[1,0],[0,1])$, where for example, $\hat{c}_{1,1}=(c_{1,1}+c_{5,1}+c_{9,1}) \bmod 2 = 1$ and $\hat{c}_{1,2}=(c_{1,2}+c_{5,2}+c_{9,2}) \bmod 2 = 0$. It is easy to check that $\hat{\mathbf{c}}$ is a codeword of $\mathbf{P}_{tb}^{(2)}$. By summing every two bits of $\hat{\mathbf{c}}$, we obtain $\hat{\mathbf{w}}=[1,1,1,1]$, a $2$-cover pseudocodeword of $C_{tb}{(2)}$. Lastly, we have $4=|\Supp(\hat{\mathbf{w}})|\leq |\Supp(\mathbf{w})|=9$, i.e., over a BEC, the pseudoweight $p(\hat{\mathbf{w}})\leq p(\mathbf{w})$. \hfill $\Box$

\end{example}

We now use Lemma \ref{tlbt} to prove our first result, that the ensemble average free pseudodistance of the unterminated SC-LDPC code ensemble can be bounded below by the pseudodistance of an associated tail-biting ensemble. 
\vspace{1mm}
\newtheorem{thm}{Theorem}
\begin{thm}[Lower bound]
The ensemble average \textit{free pseudodistance} $\bar{w}_{free}(T)$ of $E(T)$ is bounded below by $\bar{w}_{min,tb}(\lambda)$ for tail-biting termination factor $\lambda = T$, i.e.,
\begin{equation}\label{eq:lb}
     \bar{w}_{free}(T) \geq \bar{w}_{min,tb}^{(T)}.
\end{equation}
\end{thm}
\emph{Proof}. By Lemma \ref{tlbt}, for $\lambda = T$, each degree-$m$ pseudocodeword $\mathbf{w}$ for $C \in E(T)$ induces a degree-$m$ pseudocodeword $\hat{\mathbf{w}}$ for $C_{tb}(\lambda) \in E_{tb}(\lambda)$ with pseudoweight $p(\hat{\mathbf{w}})\leq p(\mathbf{w})$. Hence $w_{min,tb}^{(T)}\leq w_{free}^{(T)}$ and on average $\bar{w}_{min,tb}^{(T)} \leq \bar{w}_{free}(T)$. \hfill $\Box$

We now use the terminated ensemble to prove an upper bound on the ensemble average free pseudodistance of the unterminated SC-LDPC code ensemble.
\vspace{1mm}
\begin{thm}[Upper bound]
The ensemble average \textit{free pseudodistance} $\bar{w}_{free}(T)$ of $E(T)$ is bounded above by $\bar{w}_{min,t}(L)$ for termination factor $L = T$, i.e.,
\begin{equation}\label{eq:ub}
    \bar{w}_{free}(T) \leq \bar{w}_{min,t}^{(T)}.
\end{equation}
\end{thm}
\emph{Proof}. For every code $C=[c_1,c_2,\ldots,c_{LNb_v},\ldots]$ in $E(T)$, there corresponds a terminated code $C_t=[c_1,c_2,\ldots,c_{LNb_v}]$ in $E_{t}(L)$ with $L=T$, and every terminated code $C_t=[c_1,c_2,\ldots,c_{LNb_v}]$ in $E_{t}(L)$ with $L=T$ automatically induces a code $C=[c_1,c_2,\ldots,c_{LNb_v},0,0,\ldots]$ in $E(T)$. Consequently, for every given pair of $C$ and $C_t$, each degree-$m$ pseudocodeword of $C_t$, $\mathbf{w}_t = [w_1,w_2,\ldots,w_{LNb_v}]$, automatically induces a degree-$m$ pseudocodeword $\mathbf{w}_{[0,\infty]} = [w_1,w_2,\ldots,w_{LNb_v},0,0,\ldots]$ of $C$. Hence $w_{free}^{(T)}\leq w_{min,t}^{(T)}$ and on average $\bar{w}_{free}(T) \leq \bar{w}_{min,t}^{(T)}$. \hfill $\Box$

Without loss of clarity, we will drop the overline notation in the following discussion of ensemble average pseudodistances. 

\subsection{Free pseudodistance growth rates of SC-LDPC code ensembles}\label{sec:pdgrowth}

It has been shown in \cite{sdw11} how to calculate the asymptotic ensemble \textit{pseudoweight enumerator} for protograph-based LDPC code ensembles for a finite-degree cover. If the asymptotic pseudoweight curve has a positive zero crossing $r^+$, then it indicates that
the minimum pseudoweight typical of most members of the ensemble is close to $\delta_{min}n$ as $n\to\infty$, where $\delta_{min}$ is the \textit{pseudodistance growth rate} of the ensemble, which equals to $r^+$, and $n$ is the code length. A large pseudodistance growth rate means that, asymptotically, most pseudocodewords from the ensemble are \textit{``good pseudocodewords''}.

Similar to the definition of \emph{free distance growth rate} in \cite{mplc11}, for SC-LDPC code ensembles, we define the \emph{free pseudodistance growth rate}, $\delta_{free}^{(T)}$, to be the ratio of the free pseudodistance $w_{free}^{(T)}$ to the decoding constraint length $\nu_s$, i.e., $$\delta_{free}^{(T)}=\frac{w_{free}^{(T)}}{\nu_s}.$$ 
Then by (\ref{eq:lb}), we obtain lower bound
\begin{equation}\label{eq:grlb}
    \delta_{free}^{(T)} \geq \frac{\check{\delta}_{min}^{(T)}T}{(m_s+1)},
\end{equation}
where $\check{\delta}_{min}^{(T)} = {w_{min,tb}^{(T)}}/{n}={w_{min,tb}^{(T)}}/{(NT  b_v)}$ is the pseudodistance growth rate of $E_{tb}(\lambda)$ with $\lambda=T$ and base matrix $\btb$. Finally, by (\ref{eq:ub}), we obtain upper bound
\begin{equation}\label{eq:grub}
    \delta_{free}^{(T)} \leq \frac{\hat{\delta}_{min}^{(T)}T}{(m_s+1)},
\end{equation}
where $\hat{\delta}_{min}^{(T)} = {w_{min,t}^{(T)}}/{n}={w_{min,t}^{(T)}}/{(NT  b_v)}$ is the pseudodistance growth rate of $E_{t}(L)$ with $L=T$ and base matrix $\mathbf{B}_{[0,T-1]}$.

\subsection{Numerical results}\label{sec:num}

% Instead, we focused on improving the optimization itself. We used MOSEK \cite{mosek} as the inner optimization solver to solve the entropy maximization subproblems, the most time-consuming subroutines. For the outer optimization, we use conjugate gradient as the subproblem algorithm in Matlab \cite{matlab}. With all of these numerical tweaks and fine-tuning of the program, we are able to solve all of our numerical tests in a reasonable time without using the conjecture. 

\begin{figure}[ht]
\begin{center}
\includegraphics[width=\columnwidth]{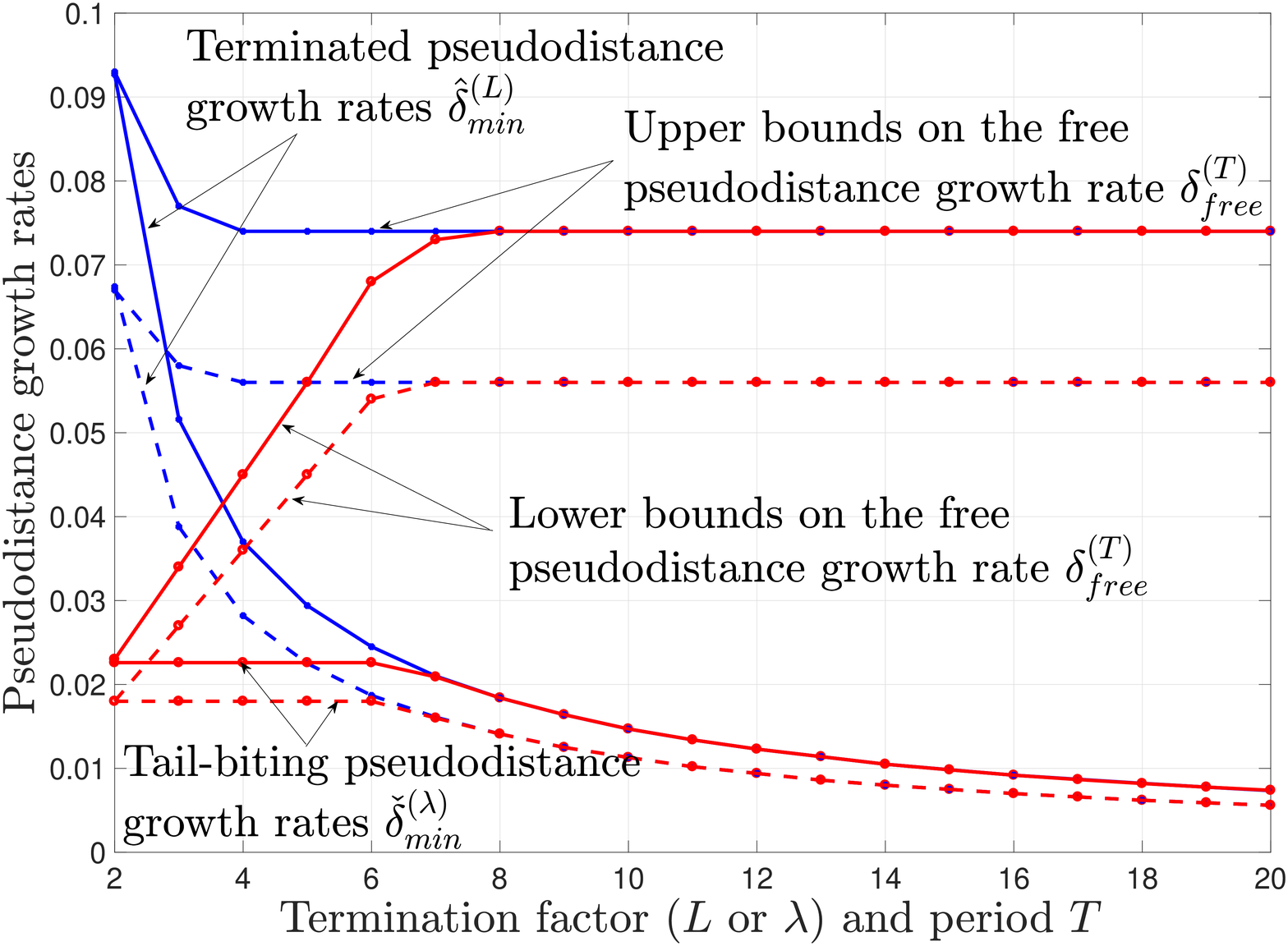}
\end{center}
\caption{Minimum pseudodistance growth rates of degree-2 covers (solid lines) and degree-3 covers (dashed lines) of terminated and tail-biting SC-LDPC code ensembles with calculated upper and lower bounds on the free pseudodistance growth rate of the associated periodically time-varying SC-LDPC code ensembles over a BEC.}\label{fig:freeps}
\end{figure}

Consider, as an example, the $(3,6)$-regular SC-LDPC code ensemble $E(T)$ with $m_s=1$ defined by (\ref{convbase}) with base matrices $\mathbf{B}_0=[\hspace{1mm}1\hspace{2mm} 2\hspace{1mm}]$ and $\mathbf{B}_1 = [\hspace{1mm}2\hspace{2mm} 1\hspace{1mm}]$. Further, consider $E^2(T)$ and $E^3(T)$, the degree-$2$ covers and degree-$3$ covers of the ensemble. Since our terminated protographs are finite, we can use the same approach from \cite{sdw11} to calculate $\check{\delta}_{min}^{(\lambda)}$ and $\hat{\delta}_{min}^{(L)}$.\footnote{Note that with our optimization framework, it was not necessary to employ the conjecture used in \cite{sdw11} to simplify the numerical calculations. We used MOSEK \cite{mosek} as the inner optimization solver to solve the entropy maximization problems, the most time-consuming subroutines. For the outer optimization, we used the conjugate gradient method as the subproblem algorithm in MATLAB.} Then, by (\ref{eq:grlb}) and (\ref{eq:grub}), we calculate the lower bound $\delta_{free}^{(T)}\geq \check{\delta}_{min}^{(T)}/2$ for $\lambda=T$ and the upper bound $\delta_{free}^{(T)}\leq \hat{\delta}_{min}^{(T)}T/2$ for $L=T$. Figure \ref{fig:freeps} shows the pseudodistance growth rate $\check{\delta}_{min}^{(\lambda)}$ (respectively, $\hat{\delta}_{min}^{(L)}$) of the tail-biting (terminated) ensembles defined by base matrix $\mathbf{B}_{tb}^{(\lambda)}$ for $\lambda=2,3,4,\ldots,20$ ($\bterm$ for $L=2,3,4,\ldots,20$) and the associated lower (upper) bound on the free pseudodistance growth rate $\delta_{free}^{(T)}$. 

%\begin{figure}[ht]
%\begin{center}
%\includegraphics[width=\columnwidth]{freeps_m3.eps}
%\end{center}
%\caption{Minimum pseudodistance growth rates of degree-3 covers of terminated and tail-biting SC-LDPC code ensembles with calculated upper and lower bounds on the free pseudodistance growth rate of the associated periodically time-varying SC-LDPC code ensembles over a BEC.}\label{fig:freeps_m3}
%\end{figure}

In Figure \ref{fig:freeps}, we observe that for degree-$2$ covers (solid lines) the tail-biting and terminated ensembles have minimum pseudoweights that grow linearly with block length, i.e., asymptotically most pseudocodewords are good. We find that the calculated tail-biting pseudodistance growth rate $\check{\delta}_{min}^{(\lambda)}$ stays constant until the termination factor $\lambda=7$ and then decreases to zero as $\lambda\to\infty$. Whereas the calculated terminated pseudodistance growth rate $\hat{\delta}_{min}^{(L)}$ decreases monotonically to zero as $L$ tends to infinity (and coincides with $\check{\delta}_{min}^{(\lambda)}$ as $L\geq 7$). More importantly, we observe that the lower and upper bounds on the free pseudodistance growth rate $\delta_{free}^{(T)}$, derived by (\ref{eq:grlb}) and (\ref{eq:grub}), coincide for $T\geq 8$, and hence gives the exact free pseudodistance growth rate, $\delta_{free}^{(T)} = 0.074$. A similar observation can be made for the degree-$3$ covers (dashed lines) in Figure \ref{fig:freeps} with exact free pseudodistance growth rate, $\delta_{free}^{(T)} = 0.056$. This implies that for degree-$2$ and degree-$3$ covers, most pseudocodewords in the unterminated SC-LDPC code ensemble are asymptotically good, and the two growth rates are significantly larger than the pseudodistance growth rates, $\delta_{min}=0.023$ and $0.018$, of the $(3,6)$-regular LDPC block code ensemble with degree-$2$ and degree-$3$ covers, respectively.

By comparing to \cite{mplc11}, we see that the exact free pseudodistance growth rate is smaller than the exact free distance growth rate, $\delta_{free}^{(T)}=0.086$. This makes sense, as explained in \cite{sdw11}, since the asymptotic ensemble pseudoweight enumerator is bounded below by the asymptotic ensemble weight enumerator, the positive zero crossing of the former is then no larger than the latter, i.e., the ensemble free pseudodistance growth rate is bounded above by the ensemble free distance growth rate. Although here the free pseudodistance growth rate is only calculated for the degree-$2$ and degree-$3$ covers of the ensemble, it is already a better indicator of the iterative decoding performance than the classical free distance growth rate. Lastly, it was observed in \cite{sdw11} that the ensemble pseudodistance growth rate decreases as the pseudocodeword cover degree $m$ increases. We see that the ensemble free pseudodistance growth rate also decreases as the pseudocodeword cover degree increases.

% \begin{figure}[h]
% \begin{center}
% \includegraphics[width=\columnwidth]{36pw.eps}
% \end{center}
% \caption{Asymptotic pseudoweight enumerators over AWGNC for the $(3,6)$ LDPC code ensemble with degree-$1$, degree-$2$, and degree-$3$ covers}\label{fig:36pw}
% \end{figure}

% As a side note, we want to point out that during our numerical experiments, we obtained different numerical results from \cite{sdw11} for the asymptotic pseudoweight enumerator of the $(3,6)$ regular LDPC code ensemble for degree-$2$ and degree-$3$ covers. Figure \ref{fig:36pw} shows the asymptotic pseudoweight enumerators over AWGNC for the $(3,6)$ LDPC code ensemble with degree-$1$, degree-$2$ and degree-$3$ covers. Note that degree-$1$ cover is nothing but the regular LDPC code itself. Our results shows that for the $(3,6)$ regular LDPC code ensemble, the minimum pseudoweight (weight if $m=1$) growth rate $\delta_{min}$ decreases monotonically as the cover degree $m$ increases, which is different from the results shown in \cite{sdw11} that the growth rates stay the same for $m=1,2,3$. Also note that for the $(3,6)$ SC-LDPC ensemble we considered, the tail-biting ensemble is equivalent to the $(3,6)$ regular LDPC code ensemble when the termination factor $\lambda = 3$. When $\lambda=3$, we observed similar results as Figure \ref{fig:36pw} over a BEC. 

\vspace{2mm}
\section{Conclusions}\label{sec:conc}
In this paper we considered pseudocodewords of periodically time-varying SC-LDPC code ensembles with finite-degree covers over a BEC. We proved that if the typical pseudodistance of the tail-biting/terminated SC-LDPC code ensemble grows linearly with the block length as the block length tends to infinity, then the typical free pseudodistance of the unterminated SC-LDPC code ensemble grows linearly as the constraint length tends to infinity. This result follows from the fact that the ensemble average minimum pseudoweight can be bounded from below (above) by the associated tail-biting (terminated) ensemble average minimum pseudoweight. We numerically evaluated the upper and lower bounds of the free pseudodistance growth rate for a $(3,6)$-regular ensemble of periodically time-varying SC-LDPC codes and found that the two bounds coincide as the period becomes sufficiently large and gives the exact free pseudodistance growth rate for the code ensemble considered. Moreover, the free pseudodistance growth rate is significantly larger than the underlying LDPC block code pseudodistance growth rate for the degree-$2$ and degree-$3$ covers considered. 

% We showed that for certain $(J,K)$-regular SC-LDPC code ensembles and a fixed cover degree, the typical pseudodistance of the unterminated (tail-biting/terminated) SC-LDPC code ensembles grows linearly with the constraint (block) length as the constraint (block) length tends to infinity. We proved that the ensemble average minimum pseudoweight can be bounded from below (above) by the associated tail-biting (terminated) ensemble average minimum pseudoweight. We then derived the lower and upper bounds for the free pseudodistance growth rate. We found that the two bounds coincide as the period becomes sufficiently large and gives the exact free pseudodistance growth rate for the code ensemble considered. We observed empirically that the free pseudodistance growth rate is significantly larger than the LDPC block code pseudodistance growth rate of degree-$2$ covers considered. 
\vspace{3mm}
\section*{Acknowledgment}
This material is based upon work supported by the National Science Foundation under Grant No. ECCS-1710920.

\vspace{5mm}
\bibliographystyle{IEEEtran}

\end{document}